\documentclass[aps,prl,groupedaddress,amsmath,amssymb]{revtex4}
\usepackage{subfigure}
\usepackage{amsfonts}
\usepackage{amsmath}
\usepackage{amssymb}
\usepackage{subfigure}
\usepackage{soul}
\usepackage{xcolor,soul}
\usepackage[utf8]{inputenc}
\usepackage{amssymb,amsthm}
\usepackage{amsmath}
\usepackage{graphicx,color}
\usepackage{bm}

\newcommand{\im}[0]{\mathrm{i}}

\begin{document}

\title{Implementation of Optimal Thermal Radiation Pumps using Adiabatically Modulated Photonic Cavities}

\author{Lucas J. Fern\'andez-Alc\'azar$^{*,1,2}$, Huanan Li$^{3}$, Mona Nafari$^{1}$ Tsampikos Kottos$^{1}$} 
\affiliation{$^1$Wave Transport in Complex Systems Lab, Department of Physics, Wesleyan University, 
Middletown, CT-06459, USA\\
$^2$Instituto de Modelado e Innovaci\'on Tecnol\'ogica (CONICET-UNNE) and Facultad de Ciencias Exactas, Naturales y Agrimensura,
Universidad Nacional del Nordeste, Avenida Libertad 5400, W3404AAS Corrientes, Argentina\\
$^3$Photonics Initiative, Advanced Science Research Center, CUNY, NY 10031, USA\\
$^*${\it email}: lfernandez@exa.unne.edu.ar}


\begin{abstract}
We numerically implement the concept of thermal radiation pumps in realistic photonic circuits and demonstrate their efficiency to control 
the radiation current, emitted between two reservoirs with equal temperature. The proposed pumping scheme involves a cyclic adiabatic 
modulation of two parameters that control the spectral characteristics of the photonic circuit. We show that the resulting pumping 
cycle exhibits maximum radiation current when a cyclic modulation of the system is properly engineered to be in the proximity of a resonance 
degeneracy in the parameter space of the photonic circuit. A developed Floquet scattering framework, which in the adiabatic limit boils down to 
the analysis of an instantaneous scattering matrix, is offering an engineering tool for designing and predicting the performance of 
such thermal pumps. Our predictions are confirmed by time-domain simulations invoking an adiabatically driven photonic cavity.  
\end{abstract}
\maketitle

\section{\label{sec:introduction} Introduction}

The subfield of thermal photonics is attracting a lot of attention due to the rapid technological progress in nano-photonics, 
and the promise that these developments can be utilized for thermal radiation management.\cite{F17,LF18,CV18,BXNKAK19}
In this respect, researchers are challenged to develop novel protocols that tackle a variety of bottlenecks imposed by fundamental 
limitations dictating thermal radiation. For example, a current research effort aims to bypass the constraints set by Kirchoff's 
and Planck's laws by capitalizing on the importance of evanescent waves in case of sub-wavelength photonic circuits.
\cite{ZF14,HSA16,GBBM18,BA16,MST16,FFFVC18a,FFFVC18b}
In parallel, other studies exploit the applicability of recent proposals for radiation control to daytime passive radiative 
cooling\cite{RRF13,RAZRF14,GS15,KJCFM17,ZMDZLTYY17}, radiative cooling of solar cells,\cite{ZRWAF14,LSCZF17} energy harvesting,
\cite{G12,L15,LPMRBA15,ZSSB16,B16,F18} thermal camouflage,\cite{LBYLQ18,K14} etc.

Along these lines, an important milestone for thermal radiation management is the implementation of photonic protocols that allow 
for a non-reciprocal heat transfer. The majority of these studies rely on magneto-optical effects\cite{A16,OMAB19a,OMAB19b} or 
non-linearities.\cite{AB13,INIT14,FTZMBBBBAMR18,KZR15} Only recently, photonic designs that enforce directional radiation 
transfer via temporal modulations are being explored. \cite{LFAESK19,BLF20,LMRA18,MA20,FAKLK21}

In this paper we consider an adiabatic thermal radiation pump which consists of a silicon-based square resonator exchanging, in the 
near-field, radiation with two thermal reservoirs which are at the same temperature. Two circular domains of the resonator are adiabatically 
driven by an out-of-phase periodic modulation. We show that, when the cyclic modulation is engineered in a way that its center, 
in the parameter space of the system, is in the proximity of a resonance degeneracy, the pumped current acquires a maximum 
value. It is interesting to point out that, on some occasions, resonance degeneracies have been associated to an exceptional point degeneracy 
of the spectrum of a (non-Hermitian) effective Hamiltonian describing the system connected to the reservoirs.\cite{LFAESK19}
Further design of the pumping cycle for our photonic pump is done following the predictions of a Floquet scattering theory, 
which, in the adiabatic limit, is utilizing an instantaneous scattering matrix (ISM), and it is confirmed via direct time-domain simulations. 

The structure of the paper is as follows. In the next section \ref{theory} we present the Floquet scattering formalism and analyze in detail 
the specific case of adiabatic pumping. Furthermore, utilizing an ISM approximation, we identify conditions which will lead to the 
design of efficient pumping circles. Based on the insights gained from the ISM Floquet formalism, in section \ref{cavity}, we design an 
adiabatic thermal radiation pump using a photonic circuit consisting of a periodically modulated microcavity. The direct time-domain 
calculations of the radiated thermal current confirm nicely the expectations, thus establishing the ISM Floquet formalism as a useful tool 
for the design of efficient pumping cycles. Our conclusions are presented at the last section of the paper.

\begin{figure}
\includegraphics[width=0.35\textwidth]{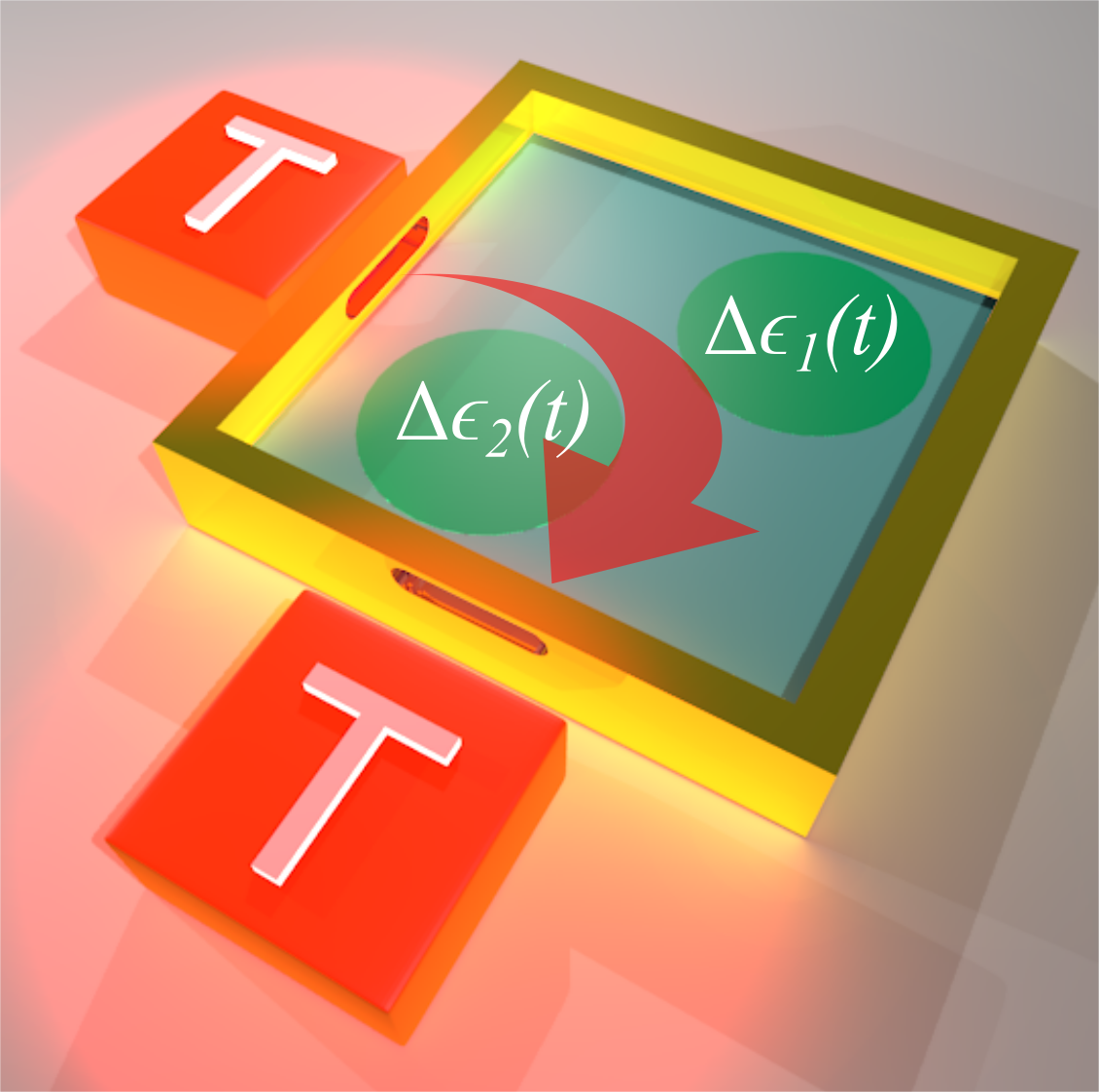} \caption{\textbf{Microcavity for adiabatic thermal radiation pumping.-} Scheme
of a two dimensional microcavity coupled to two thermal baths at the
same temperature $T$. The uniform permittivity $\varepsilon$ filling
the cavity undergoes perturbations $\Delta\varepsilon_{1}$ and $\Delta\varepsilon_{2}$
that, when periodically modulated in time, produce a nonzero current
between reservoirs. }
\label{fig1} 
\end{figure}

\section{Theoretical Considerations}\label{theory}

\subsection{Floquet scattering formalism}

We consider two thermal reservoirs $\alpha=1,2$ at temperature $T_{\alpha}$ that exchange thermal energy via a periodically 
modulated photonic circuit. The reservoirs emit radiation with amplitudes $\theta_{\alpha}^{+}(\omega)$ that satisfy the condition
\begin{equation}
\left\langle [\theta_{\alpha}^{+}(\omega)]^{*}\theta_{\beta}^{+}(\omega^{\prime})\right\rangle =\frac{1}{2\pi}{\tilde{\Theta}}_{\alpha}
(\omega)\delta_{\alpha,\beta}\delta(\omega-\omega^{\prime}),\label{correlation}
\end{equation}
where $\left\langle \cdot\right\rangle $ indicates a thermal ensemble average, and $\left\langle |\theta_{\alpha}^{+}(\omega)]|^2
\right\rangle$ describes the mean number of $\omega$-photons injected from the $\alpha-th$ thermal 
reservoir to the circuit. The mean photon number is ${\tilde{\Theta}}_{\alpha}(\omega)=\Phi_{\alpha}(\omega)\cdot\Theta_{\alpha}
(\omega)$ where $\Theta_{\alpha}(\omega)=\left(e^{\frac{\hbar\omega}{k_{B}T_{\alpha}}}-1\right)^{-1}$ and $\Phi_{\alpha}(\omega)$ 
is a spectral filter function. The later can be used to control the spectral emissivity of the thermal reservoirs and can be achieved via 
the deposition of photonic crystals that support band-gaps, or their coupling to the photonic circuit via a waveguide or a cavity with 
cut-off frequencies, etc. \cite{F17,LF18,CV18}.

We further assume that the circuit is periodically driven with a set of time-varying parameters ${\vec{x}}(t+\frac{\Omega}{2\pi})=
{\vec{x}}(t)=[x_{1}(t), \cdots,x_{N}(t)]^{T}$ which modulate, for instance, the refractive index of its constituent materials. When 
radiation at frequency $\omega$ enters into the driven circuit, it can be scattered to other sidebands with frequencies $\omega_{m}
=\omega+m\Omega,$ $m\in Z$. The incoming/outgoing radiations with amplitudes $\theta_{\alpha}^{\pm}(\omega)$ from/towards the 
reservoir $\alpha$ are related via the Floquet scattering matrix ${\bf S}^{F}$ as follows \cite{MB02,LKS18App,LSK18B} 
\begin{equation}
\theta_{\alpha}^{-}(\omega)=\sum_{\beta,m}{\bf S}_{\alpha,\beta}^{F}(\omega,\omega_{m})\theta_{\beta}^{+}(\omega_{m}),\label{theta-}
\end{equation}
where the matrix element ${\bf S}_{\alpha,\beta}^{F}(\omega_{m'},\omega_{m})$ describes the transmission/reflection amplitude of 
the electromagnetic field entering the reservoir $\alpha$ with frequency $\omega_{m'}$ provided that it was emitted from reservoir 
$\beta$ with frequency $\omega_{m}$. We furthermore assume that the number of photons is conserved during 
the Floquet scattering process,\cite{MB02} leading to a unitary Floquet scattering matrix, i.e., $({\bf S}^{F})^{\dagger}{\bf S}^{F}={\bf S}^{F}
({\bf S}^{F})^{\dagger}=I$.

We are now ready to evaluate the average photon current toward the reservoir $\alpha$. The latter takes the form
\begin{equation}
{\bar{I}}_{\alpha}^{{\rm Ph.}} = \frac{\Omega}{2\pi}\int_{0}^{\frac{2\pi}{\Omega}}dt\left[\left\langle |\theta^{-}(t)|^{2}
\right\rangle -\left\langle |\theta^{+}(t)|^{2}\right\rangle \right]
\end{equation}
where $\theta_{\alpha}^{\pm}(t)=\int_{0}^{\infty}\theta_{\alpha}^{\pm}(\omega)e^{-\im\omega t} d\omega$
is the complex field amplitude. Specifically, using Eqs.
(\ref{correlation}) and (\ref{theta-}), we obtain 
\begin{align}
{\bar{I}}_{\alpha}^{{\rm Ph.}} &= \int\frac{d\omega}{2\pi}N_{\alpha}(\omega), \label{eq:Iph} \\
N_{\alpha}(\omega) &= -{\tilde{\Theta}}_{\alpha}(\omega) + \sum_{\beta,m}|{\bf S}_{\alpha,\beta}^{F}(\omega,\omega_{m})|^{2}{\tilde{\Theta}}_{\beta}
(\omega_{m}),\notag
\end{align}
where $N_{\alpha}(\omega)$ is the photon current density entering reservoir $\alpha$. 

Similarly, we can evaluate the corresponding thermal radiation energy. The associated average energy current ${\bar{I}}_{\alpha}$ 
toward the thermal reservoir $\alpha$ reads: 
\begin{align}
{\bar{I}}_{\alpha} &= \int\frac{d\omega}{2\pi}\hbar\omega N_{\alpha}(\omega),\label{eq:I}\\
 & =\sum_{\beta,m}\int\frac{d\omega}{2\pi}\hbar\omega_{m}|{\bf S}_{\alpha,\beta}^{F}
(\omega_{m},\omega)|^{2}\left[{\tilde{\Theta}}_{\beta}(\omega)-{\tilde{\Theta}}_{\alpha}(\omega_{m})\right].\notag
\end{align}
where for the second equality we have employed the unitarity of the Floquet scattering matrix ${\bf S}^{F}$.
Equations (\ref{eq:Iph}) and (\ref{eq:I}) extend the Landauer-like formalism developed in the framework of mesoscopic quantum electronics 
\cite{MB02} to the case of photons and radiative energy currents through time modulated circuits.
 See the detailed derivation of Eqs. (\ref{eq:Iph},\ref{eq:I}) in the Supporting Information.

\subsection*{Adiabatic limit of thermal radiation}

We can do a further analytical progress with Eq. (\ref{eq:I}) in the limit of \textit{adiabatic} modulations, i.e. $\Omega\rightarrow0$,
and small temperature gradients. In 
this case, the Floquet scattering matrix ${\bf S}^{F}$ can be expressed in terms of the instantaneous scattering 
matrix $S^{t}(\omega)$ as ${\bf S}^{F}(\omega_{m},\omega) = {\bf S}^{F}(\omega,\omega_{-m})=\frac{\Omega}{2\pi}
\int_{0}^{\frac{2\pi}{\Omega}}dtS^{t}(\omega)e^{\im m\Omega t}$  (see Supporting Information). Substitution of this expression in Eq. (\ref{eq:I}) allows us to separate 
the total energy current Eq. (\ref{eq:I}) into three distinct contributions: 
\begin{equation}
\label{contributions}
{\bar{I}}_{\alpha}\approx{\bar{I}}_{\alpha,b}+{\bar{I}}_{\alpha,p}+{\bar{I}}_{\alpha,d}.
\end{equation}
The first contribution ${\bar{I}}_{\alpha,b}$ is due to the temperature gradient between the reservoirs and does not depend on the 
modulation frequency $\Omega$ 
\begin{equation}
{\bar{I}}_{\alpha,b}=\int\frac{d\omega}{2\pi}{\bar{{\cal T}}}(\omega)\hbar\omega\left[{\tilde{\Theta}}_{\alpha^{\prime}}(\omega)-
{\tilde{\Theta}}_{\alpha}(\omega)\right].
\label{Ibias}
\end{equation}
Here, ${\bar{{\cal T}}}=\frac{\Omega}{2\pi}\int dt|S_{2,1}^{t}(\omega)|^{2}$ is the average transmittance of the frozen system over one 
modulation cycle. Notice that it is the same for left and right emitted radiation since the reciprocity of the frozen (undriven) system enforces 
the relation $|S_{2,1}^{t}(\omega)|^{2}=|S_{1,2}^{t}(\omega)|^{2}$. 

The second contribution ${\bar{I}}_{\alpha,p}$ is the thermal radiative pumped energy flux, which originates from the modulation of 
the scatterer and is given as 
\begin{equation}
\bar{I}_{\alpha,p} \approx \frac{\Omega}{2\pi}\int\frac{d\omega}{2\pi}\hbar\left\{ \omega{\tilde{\Theta}}_{0}(\omega)
\frac{\partial P_{\alpha}}{\partial\omega}+
  P_{\alpha}(\omega) {\tilde{\Theta}}_{0}(\omega)\right\},\label{I_pump}
\end{equation}
where $\tilde{\Theta}_{0}(\omega)=\left.\tilde{\Theta}_{\alpha}(\omega)\right|_{T_{\alpha}=T_0}$ is associated to a mean temperature 
$T_0=(T_1+T_2)/2$, and $P_{\alpha}(\omega)=\im\int dt\left(\frac{dS^{t}}{dt}(S^{t})^{\dagger}\right)_{\alpha,\alpha}$. In the 
classical limit $\tilde{\Theta}_{0}(\omega)\approx k_{B}T_{0}/(\hbar\omega)$, the integration of 
the first term of Eq. (\ref{I_pump}) approaches to zero. We note 
that $\bar{I}_{\alpha,p}$ can be different than zero even in the absence of a temperature gradient. In addition, the 
direction of the energy current can be easily controlled by the orientation of modulation loop in the parameter space. 

Finally, the third term ${\bar{I}}_{\alpha,d}$ in Eq. (\ref{contributions}), is the analogue of a dissipation term found in adiabatic quantum 
pumps in condensed matter physics \cite{MB02,AEGS01,AEGS04,C03} and reads
 \begin{equation}
{\bar{I}}_{\alpha,d}=\frac{\Omega^{2}}{2\pi}\int\frac{d\omega}{2\pi} \hbar 
\left[\oint_{0}^{2\pi}dx\left(\frac{\partial S^{x}}{\partial x}\frac{\partial(S^{x})^{\dagger}}{\partial x}\right)_{\alpha,\alpha}\right]
\times \left[ - \frac{\partial {\tilde{\Theta}}_{\alpha}(\omega)}{\partial \omega} 
-\frac{\omega}{2}\frac{\partial^2 {\tilde{\Theta}}_{\alpha}(\omega)}{\partial \omega^2} \right],\label{Idis}
\end{equation} 
where $x=\Omega t$. In contrast to the pumped current, ${\bar{I}}_{\alpha,d}$ is proportional to ${\cal O}(\Omega^{2})$ and its 
direction does not depend on the orientation of the modulation loop, i.e., the substituting of $\Omega\rightarrow-\Omega$ does 
not reverse the direction of this current contribution. Moreover, in the adiabatic limit, the dissipative current is negligible as compared 
to the pumped current which scales as ${\cal O}(\Omega)$. Obviously, the smaller the value of ${\bar{I}}_{\alpha,d}$ the better 
is the performance of an adiabatic thermal pump.

\subsection*{Adiabatic Pumping Current}

Of particular interest is the analysis of the adiabatic energy flux between two reservoirs {\it at the same temperature}. In such 
scenario, known as adiabatic thermal radiation pumping \cite{LFAESK19}, the thermal radiative flux is dominated by the 
pumping current contribution. This can be evaluated using Eq. (\ref{I_pump}) and employing the slow modulation $\Omega
\rightarrow0$ limit. We get 
\begin{align}
{\bar{I}}_{\alpha} & \approx \frac{\Omega}{2\pi}\int\frac{d\omega}{2\pi}\hbar\omega{\tilde{\Theta}}_{0}(\omega)
{\cal A}Q_{\alpha}(\omega),\notag\\
Q_{\alpha}(\omega) & = \lim_{{\cal A}\rightarrow0^{+}}\frac{1}{{\cal A}}\left\{ \frac{1}{\omega}\frac{\partial
\left[\omega P_{\alpha}(\omega)\right]}{\partial\omega}\right\},
\label{eq:IQ}
\end{align}
where $Q_{\alpha}(\omega)$ represents the radiative energy density per pumping area and we have also assumed that the 
modulation cycle encloses a small area in the parameter space i.e., ${\cal A}=\oint\Pi_{\nu=1}^{N}dX_{\nu}\rightarrow0$. In 
the derivation of Eq. (\ref{eq:IQ}) we have also omitted the dissipative term $I_{\alpha,d}$ since it is proportional to ${\cal O}
(\Omega^2)$. 

It is interesting to note that $P_{\alpha}$ (and thus $Q_{\alpha}$) can be given conveniently in terms of the instantaneous 
scattering matrix $S^{t}$. To better understand this, we parametrize $S^{t}$ as
\begin{align}
S^{t} & =e^{\im\varphi^{t}}\begin{bmatrix}\sqrt{R^{t}}e^{\im\alpha^{t}} & \im\sqrt{1-R^{t}}\\
\im\sqrt{1-R^{t}} & \sqrt{R^{t}}e^{-\im\alpha^{t}}
\end{bmatrix},\quad0\leq R^{t}\leq1,\label{eq:Sparam}
\end{align}
where $\varphi^{t}$ is the transmission phase, $R^{t}$ is the reflectance, and $\alpha^{t}$ is the phase of the reflection 
coefficient. With this parametrization, we can show that 
\begin{equation}
\label{Pfunction}
P_{1}=-P_{2}=\int_{0}^{2\pi/\Omega}dtR^{t}\frac{d\alpha^{t}}{dt},
\end{equation}
which highlights the geometric character of the pumping operation.\cite{BTTOFA20,TWFV20}
Eq. \ref{Pfunction} also constitutes a dramatic computational simplification for the evaluation of ${\bar I}_{\alpha}$ (compare with Eqs. 
(\ref{currentTD1},\ref{currentTD}) below).

On the other hand, the radiative (time-averaged) pumped thermal energy flux ${\bar{I}}_{\alpha}$ during one pumping 
cycle can be evaluated directly from direct time domain experiments using the following expression\cite{LFAESK19}
\begin{equation}
{\bar{I}}_{\alpha} \equiv \frac{\Omega}{2\pi}\int_{0}^{2\pi/\Omega}dt\int\frac{d\omega}{2\pi}{\cal I}_{x_{0}}(t,\omega)\hbar\omega
{\tilde \Theta}_0(\omega),
\label{currentTD1}
\end{equation}
where ${\cal I}_{x_{0}}$ is the normalized time-dependent directional net energy current evaluated at some observation 
position $x_{0}$ between the system and the reservoir. Comparison between Eqs. (\ref{eq:IQ},\ref{currentTD1}) allow us to 
identify the radiative energy density (i.e. per area in the parameter space) $Q(\omega)$ to be equal to 
\begin{equation}
Q(\omega) \equiv \lim_{{\cal A}\rightarrow0}\frac{1}{{\cal A}}\int_{0}^{2\pi/\Omega}dt{\cal I}_{x_{0}}(t,\omega),
\label{currentTD}
\end{equation}
Clearly, the evaluation of the pumping current via Eq. (\ref{currentTD}) requires the knowledge of ${\cal I}_{x_{0}}(t,\omega)$ 
during the whole period $t\in[0,2\pi/\Omega)$ and for all emission frequencies $\omega$. This is a heavy computational effort 
in case of adiabatic drivings. For its calculation, we will consider uncorrelated thermal emissions at frequency $\omega$ of unit 
incident flux from the two reservoirs. In our realistic example below with an adiabatically modulated resonator, we shall utilize 
both expressions and confirm their equivalence.

\subsection*{Design Rules for Efficient Pumping Circles}

A careful analysis of the integrand in Eq. (\ref{eq:IQ}) allows us to derive a number of design rules for the engineering of 
efficient adiabatic thermal radiation pumps. First observation stems from the fact that an appropriate design of the spectral 
filtering function $\Phi(\omega)$ can lead to an enhancement/suppression of specific frequency components of $Q(\omega)$ 
thus controlling the integral in Eq. (\ref{eq:IQ}) and therefore the value of ${\bar I}_{\alpha}$. At the same time, we observe 
that even in the absence of spectral filtering (i.e.  $\Phi_{\alpha}(\omega)=1$), the smooth positive function $\omega{\tilde{
\Theta}}_{0}(\omega)$ introduces a frequency-weight that suppresses the high-frequency components of $Q_{\alpha}$ while 
enhances its corresponding low-frequency ones. Therefore, knowledge of $Q(\omega)$ through the ISM approximation may 
guide the design of efficient pumping circles. 

Another observation is associated with the dependence of $P(\omega)$ from the scattering parameters, see Eq. (\ref{Pfunction}). 
This dependence allows us to conclude that the main contribution to $\bar{\mathcal I}$ originates from a frequency range around 
transmission resonances, where $P\left(\omega\right)$ becomes significant due to the rapid changes 
of the instantaneous reflection phase $\alpha^{t}$ and reflectance $R^t$. It is natural to assume that when a pair of nearby 
resonances approach one-another, they can further enchance the sensitivity of $\alpha^t$ and $R^t$ on the parameters of the 
pump, thus increasing the pumping-induced thermal energy flux density $\bar{\cal I}$. These observations 
will be allowing us to understand and engineer the radiation current of a realistic photonic pump as we will show in the next 
section. 

\section{Microcavity implementation}\label{cavity} 

We are now ready to evaluate the thermal energy flux for a realistic photonic adiabatic pump consisting of a square optical 
microcavity, see Fig. \ref{fig1}. The cavity has linear length $a=50\mu m$ and it is formed by a dielectric material (silicon) with 
relative permittivity $\varepsilon=11.6$ (silicon). We assume that the cavity is surrounded with perfect electric conductors 
(PECs). The cavity is brought in the proximity of two reservoirs with the same temperature $T=300$ K. 

We adiabatically drive the cavity by modulating the permittivities $\varepsilon_{1(2)}(t)$ within two embedded circular 
spots while the permeability $\mu_{0}$ remains constant, see Fig. \ref{fig1}. In our simulations below, we 
will consider the specific modulation protocol $\varepsilon_{1(2)}=\varepsilon(U+\Delta\varepsilon_{1(2)})$
where $(\Delta\varepsilon_{1},\Delta\varepsilon_{2})=\frac{\Delta}{\varepsilon}(-\sin \Omega t,\cos \Omega t)$
and the offset $U$ shifts the driving circle along the line $\varepsilon_{1}=\varepsilon_{2}=
\varepsilon U$ in the parameter space $(\Delta\varepsilon_{1},\Delta\varepsilon_{2})$. This type of time-modulation of the dielectric constant of 
silicon could be achieved via carrier injection with frequency up to several GHz \cite{RMGT10}. Our goal is to design 
efficient pumping circles using the ISM Floquet formalism and confirm their efficiency by evaluating the pumped current using 
direct time-domain simulations.

\subsection*{Eigenmode Analysis}

Before analyzing the pumped energy current, we will first describe the spectral properties of the microcavity and the parametric
change of the eigenmodes with respect to the permittivity variations $\varepsilon(x,y)=\varepsilon[1+\Delta\varepsilon(x,y)]$. To 
this end, we first consider the uniform square cavity of size $a$ by setting $\Delta\varepsilon(x,y)=0$. Due to its geometric symmetry, 
the microcavity presents spectral degeneracies which we will be using for the design of optimal pumping circles. 

The spectrum can be evaluated explicitly by solving directly the frequency-domain Maxwell's equations
\begin{equation}
\nabla\times\vec{E}=-\im\omega\mu_{0}\vec{H}\ ;\ \nabla\times\vec{H}=-\im\omega\varepsilon_{0}\vec{E}.\label{Max2}
\end{equation}
We assume transverse electric (TE) modes of the form $\vec{E}=\vec{E}_{t}(x,y)$, $\vec{H}=\vec{H}_{t}(x,y)+H_{z}(x,y)\hat{z}$, 
where the vectors $\vec{E}_{t}(x,y)$ and $\vec{H}_{t}(x,y)$ with the subscript $t$ stands for transverse field components in the 
x-y plane. From Eq. (\ref{Max2}), we have $\vec{H}_{t}=0$ and $\vec{E}_{t}=(\im\omega\varepsilon)^{-1}\nabla_{t}H_{z}\times\hat{z}$,
leading to the 2D Helmholtz equation 
\begin{equation}
\nabla_{t}^{2}H_{z}(x,y)+\omega^{2}\varepsilon\mu_{0}H_{z}(x,y)=0
\label{wave}
\end{equation}
which when supplemented with the PEC boundary condition, i.e. $\frac{\partial H_{z}}{\partial n}=0$ yields 
\begin{equation}
H_{z}^{mn}=H_{mn}\cos(k_{x}x)\cos(k_{y}y),\label{sol1}
\end{equation}
where $k_{x}=m\pi/a$, $k_{y}=n\pi/a$ with $n,m=0,1,2,\cdots$. The corresponding frequency spectrum is 
\begin{equation}
\omega_{mn}=\frac{\pi}{a\sqrt{\varepsilon\mu_{0}}}\sqrt{m^{2}+n^{2}}\,\quad n,m=0,1,2,\cdots,
\end{equation}
where $\varepsilon=\varepsilon_{0}\varepsilon_{r}$ is the permitivity, and $a$ is the linear size of the square cavity. 

We focus on the two degenerate TE modes with the lowest frequencies $\omega_d/(2\pi)=\omega_{10}/(2\pi)=\omega_{01}/(2\pi)=
(2a\sqrt{\varepsilon\mu_{0}})^{-1}\approx0.8802$ THz, and the corresponding electric field profiles 
\begin{align}
\vec{E}^{01}&=E_{x}^{01}\hat{x}\ ;\ E_{x}^{01}=\im\sqrt{\frac{\mu_{0}}{\varepsilon}}\sqrt{\frac{2}{a}}\sin\left(\frac{\pi}{a}y\right)\notag\\
\vec{E}^{10}&=E_{y}^{10}\hat{y}\ ;\ E_{y}^{01}=-\im\sqrt{\frac{\mu_{0}}{\varepsilon}}\sqrt{\frac{2}{a}}\sin\left(\frac{\pi}{a}x\right).
\label{Emodes}
\end{align}
The modification of the permittivities $\varepsilon_{1(2)}$ lifts the degeneracy. Accordingly, the mode profiles will be modified and their 
electric fields will satisfy the following relation
\begin{equation}
\nabla\times\nabla\times\vec{E}=(\omega/c)^{2}(1+\Delta\varepsilon)\vec{E}.\label{eq: modified}
\end{equation}
We approximate the solution near $\omega_{d}$ via the linear combination $\vec{E}\approx\varphi_{1}\vec{E}^{01}+\varphi_{2}\vec{E}^{10}$,
and obtain the following relation using Eq. \eqref{eq: modified}
\begin{align}
\left(\frac{\omega_{d}^{2}-\omega^{2}}{c^{2}}\right)\left(\varphi_{1}\vec{E}^{01}+\varphi_{2}\vec{E}^{10}\right) & =\left(\frac{\omega}{c}\right)^{2}\Delta\varepsilon\left(\varphi_{1}\vec{E}^{01}+\varphi_{2}\vec{E}^{10}\right),
\end{align}
which, after taking the dot product with the vectors $\vec{E}^{{01}^{*}}$
and $\vec{E}^{{10}^{*}}$ and the subsequent integral over space,
yields
\begin{align}
\left(\frac{\omega_{d}^{2}-\omega^{2}}{\omega^{2}}\right)\begin{pmatrix}D^{0} & 0\\
0 & D^{0}
\end{pmatrix}\begin{pmatrix}\varphi_{1}\\
\varphi_{2}
\end{pmatrix} & =\begin{pmatrix}N^{0} & N^{01}\\
N^{10} & N^{1}
\end{pmatrix}\begin{pmatrix}\varphi_{1}\\
\varphi_{2}
\end{pmatrix}\label{eq: matrix}
\end{align}
where 
 $D^{0} =\iint \left(\vec{E}^{01}\right)^{*}\cdot\vec{E}^{01}dxdy=\iint \left(\vec{E}^{10}\right)^{*}\cdot\vec{E}^{10}dxdy$,
 $N^{0} =\iint \left|\vec{E}^{01}\right|^2\Delta\varepsilon dxdy$,  
 $N^{1} =\iint \left|\vec{E}^{10}\right|^2\Delta\varepsilon dxdy$, and
 $N^{10}=N^{{01}^{*}}=\int\vec{E}^{{10}^{*}}\cdot\vec{E}^{01}\Delta\varepsilon dxdy$.

The existence of the nonzero solution of Eq. \eqref{eq: matrix} allow us to find the frequencies of the perturbed modes: 
\begin{equation}
\frac{\Delta\omega}{\omega_{d}}\approx-\frac{N^{0}+N^{1}\pm\sqrt{\left(N^{0}-N^{1}\right)^{2}+4N^{01}N^{10}}}{4D^{0}},
\label{freq}
\end{equation}
where $\Delta\omega=\omega-\omega_{d}$. It is important to notice that $\vec{E}^{01}$ is perpendicular to $\vec{E}^{10}$ 
which results in $N^{10}=(N^{01})^{*}=0$. Consequently, the splitting of the eigenfrequencies $\Delta\omega$ would vanish 
when $N^{0}=N^{1}$, producing a {\it degenerate line} in the parameter space. 

The condition $N^{0}=N^{1}$ for degeneracy is not related to the symmetry of the system. For example, consider the case 
of a material perturbation on two separate domains with $\Delta\varepsilon(x,y)=\Delta\varepsilon_{1}(x,y)+
\Delta\varepsilon_{2}(x,y)$. In this case, we have 
\begin{equation}
N^{0}-N^{1} = \frac{\mu_{0}}{\varepsilon}\frac{2}{a}\iint dxdy \left[\sin^{2}\left(\frac{\pi}{a}x\right)+
\sin^{2}\left(\frac{\pi}{a}y \right)\right] \cdot \left[\Delta\varepsilon_{1}(x,y)+\Delta\varepsilon_{2}(x,y)\right].
\end{equation}
Therefore, for a given $\Delta\varepsilon_{1}(x,y)$, we can easily find a $\Delta\varepsilon_{2}(x,y)$ such that $N^{0}-N^{1}=0$.
In such case the degeneracy will survive even in the presence of these specific permittivity perturbations which evidently 
destroy the geometric symmetry of the cavity. Of course, the above claim is correct up to first order perturbation. It is therefore 
imperative to analyze numerically their existence and even question it in the case that the cavity is coupled to the reservoirs.
This latter situation needs additional care since the coupling to the reservoirs introduces an additional perturbation which might 
destroy the degeneracy of the associated resonant modes. 

\begin{figure}
\centering \includegraphics[width=0.5\textwidth]{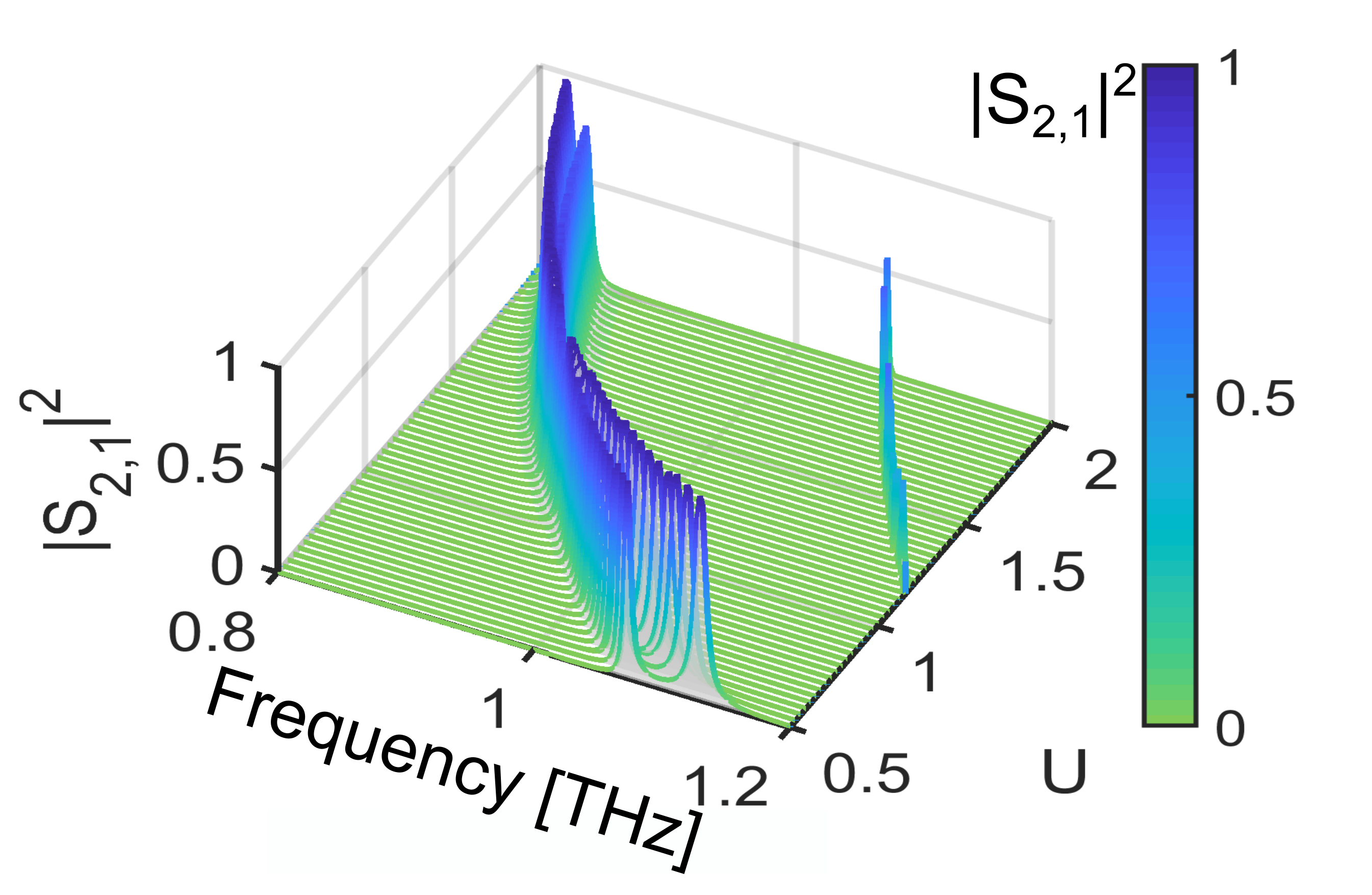}
\caption{Transmittance spectrum $|S_{2,1}|^{2}$ of the cavity (in absence
of driving) as a function of frequency $\omega/(2\pi)$ and $U$.
The peaks of the transmittance show a resonance degeneracy around
$U=1.05$.} 
\label{figSpec} 
\end{figure}

To confirm the conclusions of the perturbation argument, we analyze in detail the transmission spectrum of the cavity in a frequency 
range containing the two lowest resonances, see Fig. \ref{figSpec}. Due to the coupling to the reservoirs, the resonant frequencies 
are detuned with respect to the eigenfrequencies of the isolated resonator, leading to a destruction of the degenerate line. The 
position of the resonant modes as a function of the offset parameter $U$ is indicated in Fig. \ref{figSpec} with solid and dashed black 
lines. We are able to identify a degenerate point where the resonances coalesce as a function of the 
shift parameter $U$. The degeneracy, occurring at $\omega_d/(2\pi)\sim 0.95$THz for an offset value $U_d=1.05$, will be the starting point 
for the design of an efficient pumping circle.

\subsection*{Pumping Cycles and Optimal Thermal Radiation Currents}

\begin{figure}
\centering \includegraphics[width=0.5\textwidth]{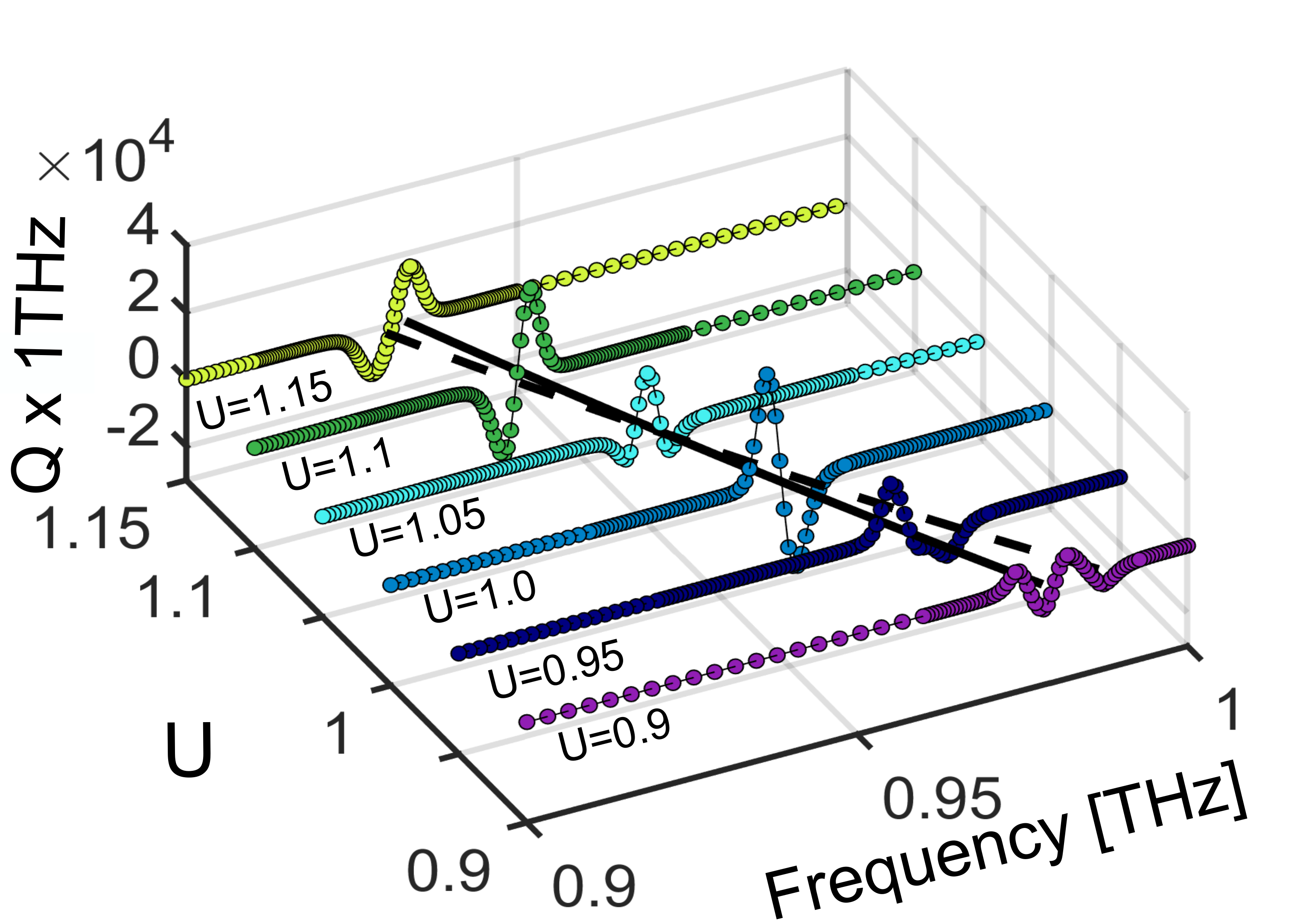} 
\caption{Radiative pumped energy density $Q(\omega)$ versus
frequency $\omega/(2\pi)$ for different values of $U$. The driving
protocol is given by the parametric cycle $\varepsilon_{1(2)}=\varepsilon(U+\Delta\varepsilon_{1(2)})$
where $(\Delta\varepsilon_{1},\Delta\varepsilon_{2})=\frac{\Delta}{\varepsilon}(-\sin x,\cos x)$
for $x\in[0,2\pi)$. The resonant frequencies, fitted from the peaks
of $|S_{2,1}|^{2}$, are shown with black solid and dashed lines.
The highest values of $Q(\omega)$ are obtained at $U=1$ and $U=1.1$
which are in the neighborhood of the resonance degeneracy. For the
calculation of $Q$ we used a parametric area ${\cal A}=\pi(\Delta/\varepsilon)^{2}$,
$\Delta=0.2$\cite{note1}, $\varepsilon=11.6$, and frequency range $[\omega_{{\rm min}},\omega_{{\rm max}})/(2\pi)=[0.88,1.06)$THz. }
\label{figQ} 
\end{figure}

We assume that the two thermal reservoirs that are attached to the microcavity are at fixed temperature $T=300$ K. We 
model these two reservoirs as two identical lumped 
ports with characteristic impedance $Z_{0}=50$ Ohms and length $\frac{1}{20}a$ (for the specific arrangement see Fig. 
\ref{fig1} and Supporting Information). The ports carry time-dependent voltage excitations originating from synthesized noise sources $V_{\alpha}$, 
which satisfy the correlation relation $\langle V_{\alpha'}(\omega)V_{\alpha}^{*}(\omega^{\prime})\rangle=\frac{2Z_{0}}{\pi}
\hbar\omega\Phi(\omega)\Theta_{C}(\omega)\delta(\omega-\omega^{\prime})\delta_{\alpha,\alpha'}$ with $\Theta_{C}
(\omega)=k_{B}T/(\hbar\omega)$.

In order to isolate the frequency domain where the resonance degeneracy $\omega_d/(2\pi)\approx 0.95THz$ occurs, we 
have filtered the emission of the baths using a spectral filtering function $\Phi(\omega)=
\theta_{H}(\omega-\omega_{{\rm min}})-\theta_{H}(\omega-\omega_{{\rm max}})$, where $\theta_{H}(\omega)$ is the Heaviside 
step function, $\omega_{\rm min}/(2\pi)=0.88THz$, and $\omega_{\rm max}/(2\pi)\le1.06THz$ was chosen in a way that maximizes 
the radiative current.

Next, we calculate the radiative energy density $Q(\omega)$ versus frequency. Although the evaluation of $Q(\omega)$ is a 
demanding computational task in case of a direct implementation of Eq. (\ref{currentTD}), the simplicity of the ISM approach 
Eq. (\ref{eq:IQ}) allows us to perform this task easily for a large number of $U$-values. The results of our calculation is shown 
in Fig. \ref{figQ}. We find that $Q(\omega)$ is significantly enhanced at frequencies $\omega$ which are at the vicinity of the 
resonance frequencies (indicated with solid and dashed black lines in Fig. \ref{figQ}). Importantly, this enhancement turns 
out to be more pronounced in the neighborhood of the resonance degeneracy around $U_d\approx1.05$. In this case, 
$Q(\omega)$ becomes symmetric around $\omega_d$ where it acquires its maximum magnitude. 
For slightly different $U$-values e.g. $U=U_d\pm 0.05$, the $Q(\omega;U)$ becomes anti-symmetric (see Fig. \ref{figQ}), 
developing two resonant peaks symmetrically placed around a particular frequency $\omega_c$ where $Q(\omega_c;U)=0$.

\begin{figure}
\centering \includegraphics[width=0.5\textwidth]{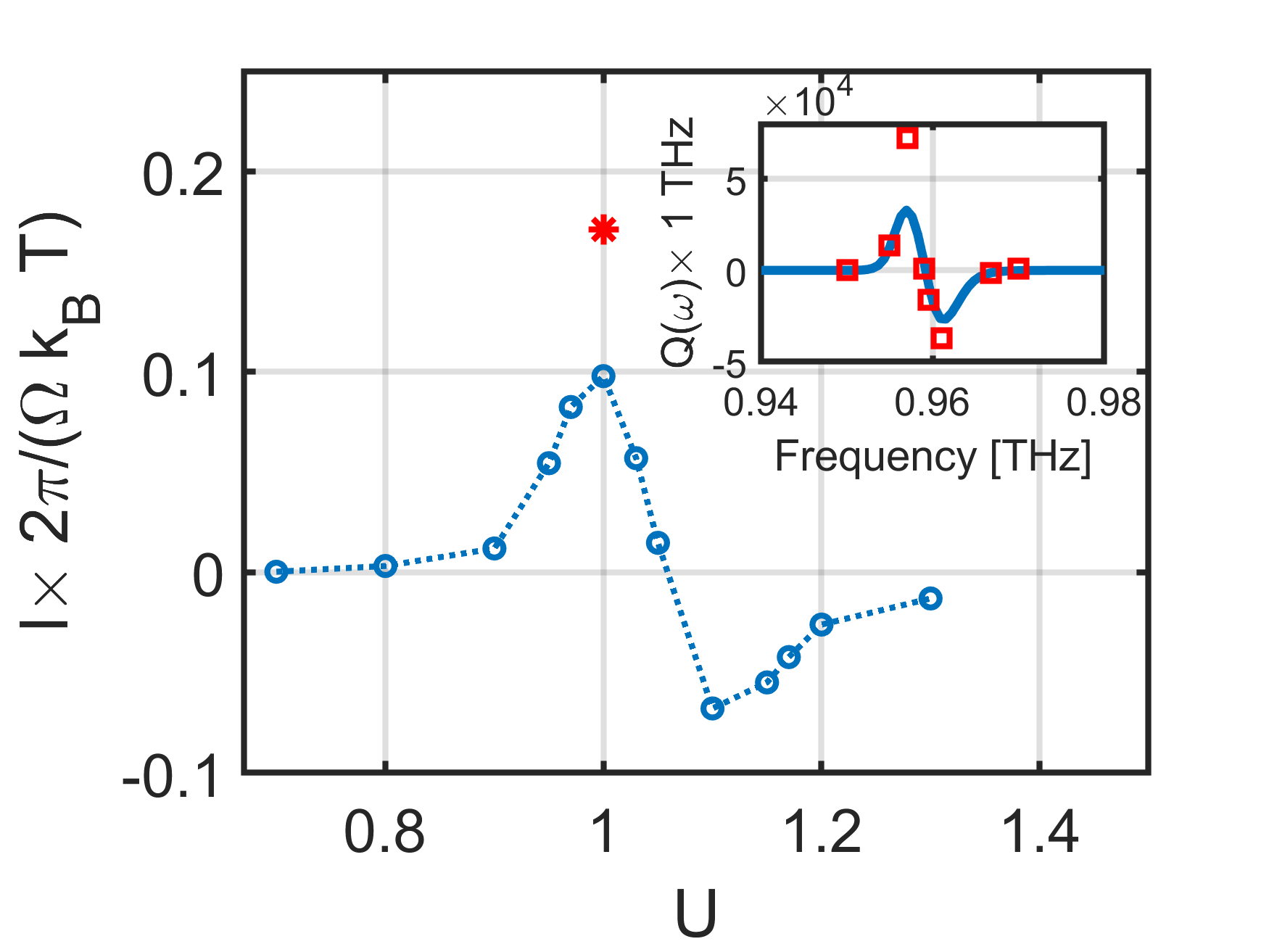} 
\caption{ Total (averaged) pumped radiative
energy current ${\bar{I}}$ versus $U$ from the integration in frequency
of $Q(\omega)$ according to Eq. (\ref{eq:IQ}), blue circles joined
through dotted line. The red asterisk indicates the associated
radiative energy current from the time domain simulations.
(Inset) Average pumped radiative energy density $Q(\omega)$ versus
frequency $\omega/(2\pi)$ for $U=1$. The solid blue line is obtained
from the scattering parameters through of Eq. (\ref{eq:IQ}) and the
red square dots are the results of the time domain simulation. For
the later we used a modulation frequency $\Omega=0.0125\omega_{0}$
where $\omega_{0}\approx2\pi\times1THz$. We set the total time to
$0.5ns$ which contains at least 5 modulation cycles, ensuring a fairly
reasonable steady state.}
\label{fig4} 
\end{figure}

The integration of $Q(\omega)$ with the weight function $\hbar \omega \Theta_{\alpha}(\omega)$ (we assume $\Phi(\omega)=1$) 
leads to a slightly enhanced contribution 
of the smaller frequencies in the radiative pumped current ${\bar I}_{\alpha}$, Eq. (\ref{eq:IQ}). However, the 
anti-symmetric shape of $Q(\omega)$ affects the calculation for ${\bar I}_{\alpha}$ such that it turns out to be (almost) zero.
Off course, one can further enhance the magnitude of the radiative current by appropriately engineering the spectral 
filtering function that applies at the reservoirs. In Fig. \ref{fig4},
we report the thermal radiation current ${\bar I}_{\alpha}$ versus the control parameter $U$ for an example case 
of a spectral filter such that $\omega_{\rm max} =\omega_d +a\cdot (U-U_d)$, 
with $a\approx-1.84 \cdot 10^{11}$, that fits the resonances around $U\approx U_d$. Such filtering, enforces a complete suppression 
of the large $\omega$-contributions to the integral of Eq. (\ref{eq:IQ}) leading to a dramatic increase of the radiation current. 
Since the radiative energy density is also anti-symmetric with respect to the offset parameter $U$ around 
the degeneracy point $U_d$ the ``sign" of the first peak will change as we are moving across $U_d$. As a result the relative 
position (with respect to $U_d$) of the driving cycle in the parameter space can be used as a reconfigurable ``knob" for 
managing the direction of the pumping flux $\bar{I}$. At the same figure, we also report (red star) 
the total radiation current using a direct implementation of Eq. (\ref{currentTD1},\ref{currentTD}) for the case of $U=1$. The 
corresponding $Q(\omega)$ which has been evaluated via time-domain simulations using Eq. (\ref{currentTD}) and it is shown 
in the inset of Fig. \ref{fig4}b. For these simulations, the photonic cavity has been illuminated with uncorrelated excitations of 
unit flux via the ports, that has been generated using monochromatic voltage sources. Once the steady state is achieved, we 
integrated the average net power over one modulation cycle to obtain $Q(\omega)$ by employing Eq. (\ref{currentTD}). 
The nice agreement between these two methods (see blue line and red squares in the inset) confirms the efficiency of the 
instantaneous scattering matrix approach.

\section{Conclusion}

We have developed a Floquet scattering formalism that allows us to design periodic, time-modulation schemes for thermal radiation management. 
In the adiabatic limit, we arrived to analytical expressions
where the net average radiative energy current can be separated into three contributions identified as bias current, pumped energy current,
and dissipation due to the driving. All these components have been conveniently expressed in terms of the instantaneous scattering matrix. 
The resulting expressions allow for an better insight to the problem of radiation management. As an example we have used them to enhance 
the radiation current in case of an adiabatic thermal pump corresponding to the scenario of thermal reservoirs with equal temperature. We 
validate the analytical results by performing realistic time domain simulations in a photonic microcavity setup. We show that when a pumping 
cyclic is designed to operate in the proximity of a spectral degeneracy, the pumped energy flux is dramatically enhanced. Furthermore the 
direction of the radiative current can be reconfigured by displaying the position of the center of the pumping cycle with respect to the degeneracy 
point in the parameter space. Our results can be used for the study of efficient thermal radiative machines and they offer a new computational tool that allow us to design efficient driving protocols for radiation management.

\section*{Supporting Information:}
Relation between Floquet and instantaneous Scattering matrices; Detailed calculations of the thermal radiative current and its
adiabatic limit; Energy balance in periodically modulated setups; Scattering analysis of the microcavity setup.
Includes Refs. \cite{MB02, Haus, LA21,FAKK21,ZF16,COMSOL}.

\section*{Acknowledgements.}
We acknowledge partial support by an ONR Grant No. N00014-16-1-2803, by an NSF Grants No. EFMA-1641109 and by 
a grant No. 733698 from Simons Collaboration in MPS. L.J.F.-A. acknowledges partial financial  support  from  CONICET.

\newpage
\begin{center}
\section{Supporting Information: Implementation of Optimal Thermal Radiation Pumps using Adiabatically Modulated Photonic Cavities}


Lucas J. Fern\'andez-Alc\'azar$^{*,1,2}$, Huanan Li$^{3}$, Mona Nafari$^{1}$ Tsampikos Kottos$^{1}$\newline

\small{
$^1$Wave Transport in Complex Systems Lab, Department of Physics, Wesleyan University, 
Middletown, CT-06459, USA\\
$^2$Instituto de Modelado e Innovaci\'on Tecnol\'ogica (CONICET-UNNE) and Facultad de Ciencias Exactas, Naturales y Agrimensura,
Universidad Nacional del Nordeste, Avenida Libertad 5400, W3404AAS Corrientes, Argentina\\
$^3$Photonics Initiative, Advanced Science Research Center, CUNY, NY 10031, USA}

\end{center}

\setcounter{equation}{0}
\setcounter{figure}{0}
\setcounter{table}{0}
\setcounter{page}{1}
\makeatletter
\renewcommand{\theequation}{S\arabic{equation}}
\renewcommand{\thefigure}{S\arabic{figure}}
\renewcommand{\thepage}{S\arabic{page}}
\renewcommand{\bibnumfmt}[1]{[S#1]}
\renewcommand{\citenumfont}[1]{S#1}


\section*{Relation between Floquet and instantaneous Scattering matrices in the adiabatic limit.}
In this section,  we describe the relation between the Floquet scattering matrix ${\bf S}^F$ and the instantaneous scattering matrix $S^t$ in
the adiabatic limit, i.e. the case of very slow modulations.

We start by considering a system that depends on a set of parameters $\vec{x}$ and its
associated scattering matrix $S(\omega,\vec{x})$, where $\omega$ is the frequency of the incident radiation. 
Next, we consider that the parameters are modulated in time, $\vec{x}=\vec{x}(t)=\vec{x}(t+2\pi/\Omega)$, and 
that such a modulation is so slow that its time scale $\tau=2\pi/\Omega$ is much larger than any other
time scale associated with the radiation dynamics. In this {\it adiabatic } regime,
the radiation traverse the structure before there is any significant change in the parameters of the system. In consequence,
we can describe the scattering process via a scattering matrix with fixed parameters,
\begin{equation}
 \theta_{\alpha}^{-}(\omega)=\sum_\beta S^t_{\alpha,\beta}(\omega) \theta_{\beta}^{+}(\omega). \label{theta-ad1}
\end{equation}
Here, the field amplitude $\theta_{\beta}^{+}(\omega)$, incident from lead $\beta$ with frequency $\omega$, is scattered toward a lead $\alpha$ with a
transmitted/reflected amplitude $S^t_{\alpha,\beta}(\omega)= S_{\alpha,\beta}(\omega,\vec{x}(t={\rm cons.}))$, where $S^t$ denotes the {\it instantaneous} Scattering matrix. 
Here, we use the superscript $t$ to emphasize the parametric dependence of the scattering matrix on time.
Indeed, it is important to observe that $S^t(\omega)$
is the scattering matrix that is available from any numerical analysis or experiment with frozen parameters.

Next, we consider the time-dependent fields $\theta_{\alpha}^{\pm}(t)=\int_{0}^{\infty}\theta_{\alpha}^{\pm}(\omega)e^{-\im\omega t} d\omega$.
In the adiabatic limit, the field scattered toward reservoir $\alpha$ 
\begin{equation}
 \theta_{\alpha}^{-}(t)= \int_{0}^{\infty} d\omega \sum_\beta S^t_{\alpha,\beta}(\omega)  \theta_{\beta}^{+}(\omega)
 e^{-\im\omega t} , \label{theta-ad2}
\end{equation}
where we have used Eq. \ref{theta-ad1}. In contrast, in the case where the parameters $\vec{x}(t)$ are 
modulated with an arbitrary modulation frequency $\Omega$, the scattered field can be described via
the Floquet scattering matrix ${\bf S}^{F}$,
\begin{eqnarray}
 \theta_{\alpha}^{-}(t)&=& \int_{0}^{\infty}d\omega \left[  \sum_{\beta,m}{\bf S}_{\alpha,\beta}^{F}(\omega,\omega_{m}) 
 \theta_{\beta}^{+}(\omega_m)\right]e^{-\im\omega t} \notag \\
 &=&\int_{0}^{\infty} d\omega\left[ \sum_{\beta,m}{\bf S}_{\alpha,\beta}^{F}(\omega_{m},\omega) 
 e^{-\im m \Omega t}\right] 
 \theta_{\beta}^{+}(\omega)e^{-\im\omega t} .  \label{theta-F}
\end{eqnarray}
Here, to arrive to Eq. (\ref{theta-F}), we first invoked Eq. (2) of the main text and then we shifted $\omega\rightarrow\omega_{-m}$ 
followed by $-m\rightarrow m$. We also assume $\theta^{+}(\omega<0)\equiv0$, consistently with conventional coupled-mode theory 
approach \cite{SHaus,SLA21}.
Of course, in the adiabatic limit, both descriptions for the outgoing scattered field $\theta^{-}(t)$ 
must be equivalent. Therefore, by comparing Eqs. (\ref{theta-ad2}) and (\ref{theta-F}), we obtain
\begin{equation}
S^t(\omega) = \sum_{m} {\bf S}^{F}(\omega_m,\omega) e^{-\im m\Omega t} ,
\label{Sad}
\end{equation}
which is valid for  $\Omega\rightarrow 0$.

Finally, we notice that Eq. (\ref{Sad}) can be identified with the Fourier series for the instantaneous scattering matrix,
$S^t(\omega) = \sum_{m} S_m(\omega) e^{-\im m\Omega t}$,
and, hence, the corresponding Fourier coefficients $S_m(\omega)$ can be associated to the Floquet scattering matrix,
\begin{eqnarray}
 S_m(\omega)\equiv {\bf S}^{F} (\omega_m,\omega)={\bf S}^{F} (\omega,\omega_{-m}), \notag \\
 S_m(\omega)=\frac{\Omega}{2\pi} \int^{ \frac{2\pi}{\Omega} }_0 dt S^t(\omega) e^{\im m \Omega t}.
 \label{eq:St_Fourier}
\end{eqnarray}
This is an useful equation since it allow us to express the Floquet scattering matrix 
in terms of the instantaneous scattering matrix, in the adiabatic limit $\Omega\rightarrow0$.
We highlight that Eq. (\ref{eq:St_Fourier}) is also discussed in the context of electron quantum transport in driven systems. \cite{SMB02}

\section*{Average Currents}

The net average photonic current directed toward the reservoir $\alpha$,  ${\bar{I}}_{\alpha}^{{\rm Ph.}}$, is the difference between the 
average current entering the reservoir $\alpha$ from the system, ${\bar{I}}_{\alpha}^{{\rm Ph. -}}$, and the average current of photons
emitted by the reservoir $\alpha$ and directed toward the system, ${\bar{I}}_{\alpha}^{{\rm Ph. +}}$,
\begin{equation}
 {\bar{I}}_{\alpha}^{{\rm Ph.}} = {\bar{I}}_{\alpha}^{{\rm Ph. -}} -{\bar{I}}_{\alpha}^{{\rm Ph. +}}, \  \  
 {\bar{I}}_{\alpha}^{{\rm Ph. -}} = \frac{\Omega}{2\pi}\int_{0}^{\frac{2\pi}{\Omega}}dt\left\langle |\theta^{-}(t)|^{2} \right\rangle , \  \ 
 {\bar{I}}_{\alpha}^{{\rm Ph. +}}=\frac{\Omega}{2\pi}\int_{0}^{\frac{2\pi}{\Omega}}dt  \left\langle |\theta^{+}(t)|^{2}\right\rangle. 
\end{equation}
By using the Fourier transform, $\theta_{\alpha}^{\pm}(t)=\int_{0}^{\infty}\theta_{\alpha}^{\pm}(\omega)e^{-\im\omega t} d\omega$, we obtain
\begin{equation}
 {\bar{I}}_{\alpha}^{{\rm Ph. \pm}} 
 = \frac{\Omega}{2\pi}\int_{0}^{\frac{2\pi}{\Omega}}dt \iint d\omega d\omega^\prime  
 \left\langle [\theta_\alpha^\pm(\omega)]^* \cdot \theta_\alpha^\pm(\omega^\prime) \right\rangle e^{\im (\omega-\omega^\prime)t},  \label{IPh+-}
\end{equation}
which translates the problem to the calculation of the correlation of the field amplitudes. In the case of ${\bar{I}}_{\alpha}^{{\rm Ph. +}} $, the corresponding
correlation $\left\langle [\theta_\alpha^+(\omega)]^* \cdot \theta_\alpha^+(\omega^\prime) \right\rangle$ is dictated by Eq. (1) of the main text, and thus
\begin{equation}
{\bar{I}}_{\alpha}^{{\rm Ph. +}} =\int \frac{d\omega}{2\pi} {\tilde \Theta}_\alpha(\omega). \label{IPh+} 
\end{equation}
Now, for the case of ${\bar{I}}_{\alpha}^{{\rm Ph. -}} $, the associated correlation results
\begin{eqnarray}
 \left\langle [\theta_\alpha^-(\omega)]^* \cdot \theta_\alpha^-(\omega^\prime) \right\rangle &=&
 \left\langle \left[ \sum_{\beta,m} {\bf S}^F_{\alpha,\beta}(\omega,\omega_m) \theta_\beta^+(\omega_m)\right]^* \cdot 
  \left[ \sum_{\gamma,n} {\bf S}^F_{\alpha,\gamma}(\omega^\prime,\omega^\prime_{n}) \theta_{\gamma}(\omega^\prime_{n})\right]\right\rangle, \notag \\
  &=& \sum_{\beta,m}\sum_{\gamma,n} [{\bf S}^F_{\alpha,\beta}(\omega,\omega_m)]^* {\bf S}^F_{\alpha,\gamma}(\omega^\prime,\omega^\prime_{n}) \cdot \left\langle [ \theta_\beta^{+}(\omega_m)]^* \cdot \theta_{\gamma}(\omega^\prime_{n}) \right\rangle, \label{correlation-}\\
  &=& \sum_{\beta,m}\sum_{\gamma,n} [{\bf S}^F_{\alpha,\beta}(\omega,\omega_m)]^* \cdot {\bf S}^F_{\alpha,\gamma}(\omega^\prime,\omega^\prime_{n}) \cdot \frac{1}{2\pi} \delta_{\beta,\gamma} \ \delta(\omega_m - \omega^\prime_{n}) {\tilde \Theta}_{\beta}(\omega_m), \notag
\end{eqnarray}
where we have used Eqs. (1) and (2) of the main text. Plugging Eq. (\ref{correlation-}) into Eq. (\ref{IPh+-}) we have
\begin{eqnarray}
  {\bar{I}}_{\alpha}^{{\rm Ph. -}}  &=& \iint d\omega d\omega^\prime  \sum_{\beta,m} \sum_{\gamma,n} 
  [{\bf S}^F_{\alpha,\beta}(\omega,\omega_m)]^* \cdot {\bf S}^F_{\alpha,\gamma}(\omega^\prime,\omega^\prime_{n}) \times \notag \\
  &\times& \frac{1}{2\pi} \delta_{\beta,\gamma} \ \delta(\omega_m - \omega^\prime_{n}) {\tilde \Theta}_{\beta}(\omega_m)
  \cdot \left( \frac{\Omega}{2\pi}\int_{0}^{\frac{2\pi}{\Omega}}dt e^{\im (\omega-\omega^\prime)t}  \right)
\end{eqnarray}
Here, the delta function enforces $\omega_m - \omega^\prime_{n}=0$, which implies that the difference between the central frequencies is
a multiple of the modulation frequency, i.e. $\omega-\omega^\prime=(n - m )\Omega$. In turn, this imposes an additional relation
between the integers $n,m$, because 
\begin{equation}
  \frac{\Omega}{2\pi}\int_{0}^{\frac{2\pi}{\Omega}}dt e^{\im (\omega-\omega^\prime)t}
  =\frac{\Omega}{2\pi}\int_{0}^{\frac{2\pi}{\Omega}}dt e^{\im (n-m)\Omega t}
  =\delta_{n,m}.
\end{equation}
Of course, the relation $n=m$ implies that $\omega=\omega^\prime$, and as a result
\begin{equation}
   {\bar{I}}_{\alpha}^{{\rm Ph. -}}  = \int \frac{d\omega}{2\pi} \sum_{\beta,m} |{\bf S}^F_{\alpha,\beta}(\omega,\omega_m)|^2 {\tilde \Theta}_{\beta}(\omega_m). \label{IPh-}
\end{equation}
This equation admits to be interpreted in a Landauer like formalism:
the average number of photons current directed toward reservoir $\alpha$ from the system consists of all those photons emitted at a reservoir $\beta$ with frequency $\omega_m$ that after an inelastic scattering process with the time-modulated system end up entering reservoir $\alpha$ with frequency $\omega$.
 
Finally, collecting the results of Eqs. (\ref{IPh+}) and (\ref{IPh-}), we have that the net average photon current entering reservoir $\alpha$ is
\begin{equation}
   {\bar{I}}_{\alpha}^{{\rm Ph.}}  = \int \frac{d\omega}{2\pi} \left[ \sum_{\beta,m} |{\bf S}^F_{\alpha,\beta}(\omega,\omega_m)|^2 {\tilde \Theta}_{\beta}(\omega_m) - {\tilde \Theta}_\alpha(\omega) \right].   \label{IPh_tot_N}
\end{equation}
This equation allow us to introduce the net number of photons density,
\begin{equation}
     N_\alpha(\omega)= \sum_{\beta,m} |{\bf S}^F_{\alpha,\beta}(\omega,\omega_m)|^2 {\tilde \Theta}_{\beta}(\omega_m) - {\tilde \Theta}_\alpha(\omega), \label{NPh}
\end{equation}
which is the net average number of photons per unit frequency entering reservoir $\alpha$.
Then, the net energy carried by $\omega-$photons is simply $\hbar\omega N_\alpha(\omega)$, and thus
\begin{equation}
   {\bar{I}}_{\alpha}= \int \frac{d\omega}{2\pi} \hbar \omega N_\alpha(\omega).   \label{I_tot}
\end{equation}
represents the associated thermal radiation energy current entering reservoir $\alpha$.

\section*{Adiabatic Limit for the Thermal Radiative Energy Current}
In this section we present a detailed derivation of Eqs. (5-9) of the main text. In particular, we show how, resorting to the adiabatic limit
and for small temperature gradients, we obtain analytical equations for the bias current, the pumped current, and dissipation associated 
to the driving. We highlight that all these equations are conveniently expressed in terms of the instantaneous scattering matrix.

We start by considering the thermal radiative energy current given by Eqs. (\ref{NPh}) and (\ref{I_tot}),
\begin{eqnarray}
   {\bar{I}}_{\alpha}&=& \int \frac{d\omega}{2\pi} \hbar \omega 
    \left[  \sum_{\beta,m} |{\bf S}^F_{\alpha,\beta}(\omega,\omega_m)|^2 {\tilde \Theta}_{\beta}(\omega_m) - 
    {\tilde \Theta}_\alpha(\omega) \right], \notag  \\
   &=& \int \frac{d\omega}{2\pi} \hbar \omega 
   \left[ \sum_{\beta,m} |{\bf S}^F_{\alpha,\beta}(\omega,\omega_m)|^2 {\tilde \Theta}_{\beta}(\omega_m) -
   \left( \sum_{\beta,m} |{\bf S}^F_{\alpha,\beta}(\omega,\omega_m)|^2 \right){\tilde \Theta}_\alpha(\omega) \right], \label{Iad1} \\
   &=&  \int \frac{d\omega}{2\pi} \sum_{\beta,m}  \hbar \omega_m  |{\bf S}^F_{\alpha,\beta}(\omega_m,\omega)|^2 
   \left[ {\tilde \Theta}_\beta(\omega) - {\tilde \Theta}_\alpha (\omega_m) \right],\notag
\end{eqnarray}
where we used Eq. (\ref{eq:Sunit_1}) in the second line and in the third one, we changed variables $\omega\rightarrow \omega_{-m}$
and $m\rightarrow-m$. The last line of Eq. (\ref{Iad1}) corresponds to Eq. (5) of the main text.

In the adiabatic regime, the small modulation frequency allow us to use a series expansion of the reservoir's spectral emission 
function centered in a frequency $\omega$,  
${\tilde \Theta}_\alpha (\omega_m)\approx 
{\tilde \Theta}_\alpha (\omega) +\frac{\partial {\tilde \Theta}_\alpha (\omega)}{\partial \omega}m\Omega +
\frac{1}{2}\frac{\partial^2 {\tilde \Theta}_\alpha (\omega)}{\partial \omega^2} (m\Omega)^2$.
Notice that in our case of resonant scattering, the frequency window of major contribution for the pumping process 
occurs for frequencies around some resonant frequency, i.e. $\omega\sim \omega_0>0$, which then admits the expansion 
when the modulation frequency is small. Using this approximation in Eq. (\ref{Iad1}), we obtain
\begin{eqnarray}
   {\bar{I}}_{\alpha}&\approx& \int \frac{d\omega}{2\pi} \sum_{\beta,m}  \hbar (\omega + m\Omega) |{\bf S}^F_{\alpha,\beta}(\omega_m,\omega)|^2 
   \left[ {\tilde \Theta}_\beta(\omega) - {\tilde \Theta}_\alpha (\omega) - \frac{\partial {\tilde \Theta}_\alpha (\omega)}{\partial \omega}m\Omega
   -\frac{1}{2}\frac{\partial^2 {\tilde \Theta}_\alpha (\omega)}{\partial \omega^2} (m\Omega)^2   \right],\notag \\
   &=& J_1 + J_2 + J_3 + J_4 + J_5 + {\cal O}(\Omega^3) \label{Iad_parts},
\end{eqnarray}
where 
\begin{eqnarray}
 J_1 &=& \int \frac{d\omega}{2\pi} \sum_{\beta} \hbar \omega \left[ \sum_m |{\bf S}^F_{\alpha,\beta}(\omega_m,\omega)|^2 \right] 
 \left( {\tilde \Theta}_\beta(\omega) - {\tilde \Theta}_\alpha (\omega) \right), \notag \\
 J_2 &=& \int \frac{d\omega}{2\pi} \sum_{\beta} \hbar \Omega \left[ \sum_m |{\bf S}^F_{\alpha,\beta}(\omega_m,\omega)|^2 \cdot m \right] 
 \left( {\tilde \Theta}_\beta(\omega) - {\tilde \Theta}_\alpha (\omega) \right), \notag \\
 J_3 &=& \int \frac{d\omega}{2\pi} \sum_{\beta} \hbar \omega \Omega\left[ \sum_m |{\bf S}^F_{\alpha,\beta}(\omega_m,\omega)|^2 \cdot m \right] 
  \left( -\frac{\partial {\tilde \Theta}_\alpha (\omega)}{\partial \omega} \right),  \\
 J_4 &=& \int \frac{d\omega}{2\pi} \hbar \Omega^2 \left[ \sum_{\beta,m }|{\bf S}^F_{\alpha,\beta}(\omega_m,\omega)|^2 \cdot m^2 \right] 
  \left( -\frac{\partial {\tilde \Theta}_\alpha (\omega)}{\partial \omega} \right), \notag \\
 J_5 &=& \int \frac{d\omega}{2\pi} \hbar \omega \frac{\Omega^2}{2} \left[ \sum_{\beta,m} |{\bf S}^F_{\alpha,\beta}(\omega_m,\omega)|^2 \cdot m^2 \right] 
  \left( -\frac{\partial^2 {\tilde \Theta}_\alpha (\omega)}{\partial \omega^2} \right). \notag 
\end{eqnarray}
Notice that each term depends in a different manner with $\Omega$, specifically, $J_1={\cal O}(\Omega^0)$, while $J_{2(3)}={\cal O}(\Omega^1)$ and
$J_{4(5)}={\cal O}(\Omega^2)$. 

In what follows, we discuss each one of the $J_n$ terms and their associated adiabatic limit for 
the case of systems with two reservoirs. Under such conditions, we connect the Floquet scattering matrix to the instantaneous
scattering matrix via the set of Eqs. (\ref{aux}), thus providing simple expressions for the terms in the thermal radiation current.

We start by considering the term $J_1$, which is proportional to the unbalance between the spectral emission functions 
of the  reservoirs, e.g. due to a temperature gradient (or bias). This term corresponds to the average bias thermal radiative current
discussed in Eq (7) of the main text. Here,
\begin{eqnarray}
 {\bar I}_{\alpha,b} \equiv J_1 &=&  \int \frac{d\omega}{2\pi} \sum_{\beta} \hbar \omega \bar{ {\cal T}}_{\alpha,\beta}(\omega) 
 \left( {\tilde \Theta}_\beta(\omega) - {\tilde \Theta}_\alpha (\omega) \right), \notag \\
 &=&  \int \frac{d\omega}{2\pi}  \hbar \omega \bar{ {\cal T}}(\omega) 
 \left( {\tilde \Theta}_{\alpha^\prime}(\omega) - {\tilde \Theta}_\alpha (\omega) \right) \label{SIbias}
\end{eqnarray}
where $\bar{ {\cal T}}_{\alpha,\beta}(\omega) = 
\frac{\Omega}{2\pi} \int dt |S^t_{\alpha,\beta}(\omega)|^2 
  = \frac{1}{2\pi} \int dx |S^x_{\alpha,\beta}(\omega)|^2$. 
For the second line we considered case of a system connected to two reservoirs, $\alpha\ne\alpha^\prime$, and we denote
${\bar {\cal T}}(\omega) = {\bar {\cal T}}_{\alpha,\alpha^\prime}(\omega)$.

We now consider the second and third terms, $J_2$ and $J_3$. Using Eq. (\ref{aux}) in the adiabatic limit, they read 
\begin{eqnarray}
 J_2 &=&  \int \frac{d\omega}{2\pi} \hbar \Omega
 \left[  \frac{\im}{2\pi} \int dt  \frac{\partial S^t_{\alpha,\alpha^\prime}(\omega)}{\partial t}  \left( S^t_{\alpha,\alpha^\prime}(\omega)\right)^*    \right]
 \left( {\tilde \Theta}_{\alpha^\prime}(\omega) - {\tilde \Theta}_\alpha (\omega) \right), \text{ and } \notag\\
 J_3 &=& \int \frac{d\omega}{2\pi} \hbar \omega \Omega 
 \left\{ \frac{\im}{2\pi} \int dt \left[ \frac{\partial S^t(\omega)}{\partial t}  \left( S^t(\omega)\right)^\dagger\right]_{\alpha,\alpha} \right\} 
  \left( -\frac{\partial {\tilde \Theta}_\alpha (\omega)}{\partial \omega} \right),  \\
  &=& \frac{\Omega}{2\pi} \int \frac{d\omega}{2\pi} {\tilde \Theta}_\alpha (\omega) \frac{\partial}{\partial\omega} 
  \left\{ \im \hbar\omega \int dt \left[ \frac{\partial S^t(\omega)}{\partial t}  \left( S^t(\omega)\right)^\dagger\right]_{\alpha,\alpha}, \right\} \notag
\end{eqnarray}
where we considered two reservoirs, $\alpha\ne \alpha^\prime$. To arrive to the last line, we used integration by parts
and dropped the contour terms because $\left. S^t(\omega)\right|_{\omega=0}\approx \left. S^t(\omega)\right|_{\omega\rightarrow\infty}\approx 0$.
We take the derivative in $\omega$ in $J_3$, and, for simplicity, we define
$A_{\alpha,\beta}=\im \int dt  \frac{\partial S^t_{\alpha,\beta}(\omega)}{\partial t}  \left( S^t_{\alpha,\beta}(\omega)\right)^* $,
$P_\alpha=A_{\alpha,\alpha}+A_{\alpha,\alpha^\prime}$. Both $J_2$ and $J_3$ are terms of first order in $\Omega$ so we consider their sum $J_2+J_3$. Thus,
\begin{eqnarray}
J_2 + J_3 &=& \frac{\Omega}{2\pi}  \int \frac{d\omega}{2\pi} \left\{ \hbar \omega \frac{\partial  P_\alpha }{\partial \omega} {\tilde \Theta}_\alpha (\omega) 
+ \hbar A_{\alpha,\alpha^\prime}{\tilde \Theta}_{\alpha^\prime} (\omega)+ \hbar A_{\alpha,\alpha}{\tilde \Theta}_{\alpha} (\omega) \right\}, \\
&=& \frac{\Omega}{2\pi}  \int \frac{d\omega}{2\pi} \left\{ \hbar \omega \frac{\partial  P_\alpha }{\partial \omega} {\tilde \Theta}_\alpha (\omega) 
+ \hbar P_\alpha                                     \frac{ {\tilde \Theta}_{\alpha} (\omega)+{\tilde \Theta}_{\alpha^\prime} (\omega)}{2} 
+ \hbar (A_{\alpha,\alpha}-A_{\alpha,\alpha^\prime}) \frac{ {\tilde \Theta}_{\alpha} (\omega)-{\tilde \Theta}_{\alpha^\prime} (\omega)}{2}  \right\}, \notag
\end{eqnarray}
Next, we consider the pumping scenario, where there is no temperature gradient between the reservoirs, 
$\frac{ {\tilde \Theta}_{\alpha} (\omega)+{\tilde \Theta}_{\alpha^\prime} (\omega)}{2} = {\tilde \Theta}_{0} (\omega)$,  and 
$\frac{ {\tilde \Theta}_{\alpha} (\omega)-{\tilde \Theta}_{\alpha^\prime} (\omega)}{2} = 0$.
In such a situation, $J_2+J_3$ represents the thermal radiative pumped energy current. Therefore
\begin{eqnarray}
{\bar I}_{\alpha,p}\equiv J_2 + J_3   &=& \frac{\Omega}{2\pi}  \int \frac{d\omega}{2\pi} \left\{ \hbar \omega \frac{\partial  P_\alpha }{\partial \omega} {\tilde \Theta}_{0} (\omega)
+ \hbar P_\alpha  {\tilde \Theta}_{0} (\omega)   \right\}, \label{SI_pump} \\
&=& \frac{\Omega}{2\pi}  \int \frac{d\omega}{2\pi} {\tilde \Theta}_{0} (\omega)  \hbar \frac{\partial   }{\partial \omega} \left\{  \omega P_\alpha  \right\}. \notag
\end{eqnarray}
Of course, our analysis can be extended to situations where the temperature gradient is small, where 
${\tilde \Theta}_{0} (\omega)$ is associated to a mean temperature $T_0=(T_{\alpha}+T_{\alpha^\prime})/2$.
Notice that Eq. (\ref{I_pump}) corresponds to Eqs. (8) and (10) of the main text.

Finally, we consider $J_4$ and $J_5$,
\begin{eqnarray}
 J_4 &=& 
 \frac{\Omega^2}{2\pi} \int \frac{d\omega}{2\pi} \hbar  \frac{\partial {\tilde \Theta}_\alpha (\omega)  }{\partial\omega}
  \left\{  \int dx \left[ \frac{\partial^2 S^x(\omega)}{\partial x^2}  \left( S^x(\omega)\right)^\dagger\right]_{\alpha,\alpha}  \right\}, \notag \\
J_5 &=& \frac{\Omega^2}{2\pi}  \int \frac{d\omega}{2\pi} \hbar \omega
\left\{  \int dx \left[ \frac{\partial^2 S^x(\omega)}{\partial x^2}  \left( S^x(\omega)\right)^\dagger\right]_{\alpha,\alpha} \right\} 
\frac{1}{2}\frac{\partial^2 {\tilde \Theta}_\alpha (\omega)  }{\partial\omega^2}, \label{SIdis}
\end{eqnarray}
where we made explicit the dependence of the instantaneous scattering matrix on the dimensionless parameter $x=\Omega t$ 
to emphasize the $\Omega^2$ dependence. For this reason, the sum of these two terms, ${\bar I}_{\alpha,d}\equiv  J_4 + J_5 $,
represents a dissipation current whose direction does not change by reversing the direction of the modulation cycle, 
i.e. $\Omega\rightarrow-\Omega$. Instead, it is always directed from the scatterer to the reservoir (if ${\bar I}_{\alpha,d}>0$)
and it is associated to a power exchange between the radiation and the modulation mechanism.  
In a pumping scenario, where the modulation is slow, this term is much smaller than the pumping currents and can be neglected.

Notice that ${\bar I}_{\alpha,d}\equiv  J_4 + J_5 $ constitutes Eq. (9) of the main text. This becomes evident by
using the relation $\int dx \left[ \frac{\partial^2 S^x(\omega)}{\partial x^2}  \left( S^x(\omega)\right)^\dagger\right]
=-\int dx \left[ \frac{\partial S^x(\omega)}{\partial x}   \frac{\partial \left(S^x(\omega)\right)^\dagger}{\partial x}\right]$, 
where $x\in[0,2\pi)$.


\section*{Energy balance.}
Here, we discuss about the energy balance of the time-modulated setup in the framework of the Floquet scattering. In addition, in 
the adiabatic regime, we provide an expression for the power performed on the system by the modulation mechanism.
We consider, for simplicity, the case of a system connected to two reservoirs. 

In the case where the parameters of the system are not modulated in time, the sum of the thermal radiative energy currents entering 
the reservoirs is zero, due to energy conservation. However, this is in contrast to what happens in the time modulated scenario. 
In such a situation, the sum of the thermal radiative energy currents entering the reservoirs (averaged over one modulation cycle) 
can be different than zero since the modulation mechanism may exchange energy with the radiation inside the structure. Thus,
\begin{equation}
 \sum_\alpha {\bar I}_{\alpha} = \frac{\Omega}{2\pi}W, \label{I=W}
\end{equation}
where $W$ is the average energy exchanged between the modulation mechanism and radiation per modulation cycle.
In order to find an expression for $W$ based on the radiative energy currents entering the reservoirs, we introduce
Eq. (\ref{Iad1}) into Eq. (\ref{I=W}),
\begin{eqnarray}
  \sum_\alpha {\bar I}_{\alpha} &=& \int \frac{d\omega}{2\pi} \hbar \omega 
    \left[  \sum_{\alpha,\beta,m} |{\bf S}^F_{\alpha,\beta}(\omega,\omega_m)|^2 {\tilde \Theta}_{\beta}(\omega_m) - 
    \sum_{\alpha}{\tilde \Theta}_\alpha(\omega) \right], \notag  \\
    &=& \int \frac{d\omega}{2\pi} 
    \left[  \sum_{\alpha,\beta,m} \hbar \omega_m |{\bf S}^F_{\alpha,\beta}(\omega_m,\omega)|^2 {\tilde \Theta}_{\beta}(\omega) - 
    \hbar \omega \sum_{\alpha}{\tilde \Theta}_\alpha(\omega) \right], \label{WFloquet}\\
    &=& \int \frac{d\omega}{2\pi} \sum_{\alpha,\beta,m} \hbar m\Omega|{\bf S}^F_{\alpha,\beta}(\omega_m,\omega)|^2 
    {\tilde \Theta}_{\beta}(\omega) . \notag
\end{eqnarray}
Here, to arrive to the second line, we have used the change of variables $\omega\rightarrow\omega-m$ followed by $-m\rightarrow m$ 
in the first term, and for the third line, $\omega_m=\omega+m\Omega$ and Eq. (\ref{eq:Sunit_1}).
This equation represents the power exchanged with the modulation mechanism in terms of the Floquet scattering processes inside the 
time-modulated setup. Of course, we can simplify further Eq. (\ref{WFloquet}) by considering the adiabatic limit.

In the adiabatic modulation regime, we connect the Floquet  with the instantaneous  scattering matrix via Eq. (\ref{aux}). Thus,
\begin{eqnarray}
\sum_\alpha {\bar I}_{\alpha} &=& \frac{\Omega}{2\pi} \int \frac{d\omega}{2\pi} \hbar \sum_{\beta} 
\left(\sum_\alpha A_{\alpha,\beta}(\omega)\right)  {\tilde \Theta}_{\beta}(\omega),\notag\\
  &=&  \frac{\Omega}{2\pi} \int \frac{d\omega}{2\pi} \hbar \sum_{\beta} P_\beta (\omega) \cdot  {\tilde \Theta}_{\beta}(\omega),
\end{eqnarray}
where we used $\sum_\alpha A_{\alpha,\beta}(\omega)= P_\beta (\omega)$. This last statement considers that the
instantaneous scattering matrix is reciprocal, i.e. $S^t_{\alpha,\beta}=S^t_{\beta,\alpha}$, which is indeed the case 
in the current study. Nevertheless, it is not necessary to assume reciprocity, and in such a case, 
$\sum_\alpha A_{\alpha,\beta}(\omega)=\im \int dt \left[ (S^t)^\dagger \frac{\partial S^t}{\partial t} \right]_{\beta,\beta}$.
In the simple case when the system is attached to only two reservoirs, $\alpha$ and $\alpha^\prime$, we have
\begin{equation}
 \sum_\alpha {\bar I}_{\alpha}= \frac{\Omega}{2\pi} W=\frac{\Omega}{2\pi} \int \frac{d\omega}{2\pi} \hbar P_0 (\omega) \cdot 
 \left[{\tilde \Theta}_{\alpha}(\omega)-{\tilde \Theta}_{\alpha^\prime}(\omega)\right],
\end{equation}
where $P_\alpha=-P_{\alpha^\prime}=P_0$. This indicates that the modulation mechanism has to perform a work on the system
that is used to pump radiation against the temperature gradient. 
Such an expression for the work performed on the system is analogous to the one found in the framework of
thermal motors \cite{SFAKK21}, which could be considered as the time-reverted scenario of a pump. 
There, a temperature gradient between reservoirs adiabatically drives certain degrees of freedom of the system, hence producing work.

It is interesting to note that, in the case of two reservoirs at the same temperature,
no external energy is introduced by the modulation in order to produce a thermal radiative pumping current. 
This seemingly contradictory result is, however, consistent with the picture of moving the photons between two regions within the same
energy shell. This scenario is reminiscent of situations where magnetic fields can produce a persistent radiation current 
at no energy cost \cite{SZF16}, and of course is consistent with the laws of thermodynamics.

\section*{Scattering analysis of the microcavity setup.}

In this section, we provide details associated with the calculation of the  scattering spectrum shown in Fig. 2 of the main text.

\begin{figure}
\center\includegraphics[width=0.45\textwidth]{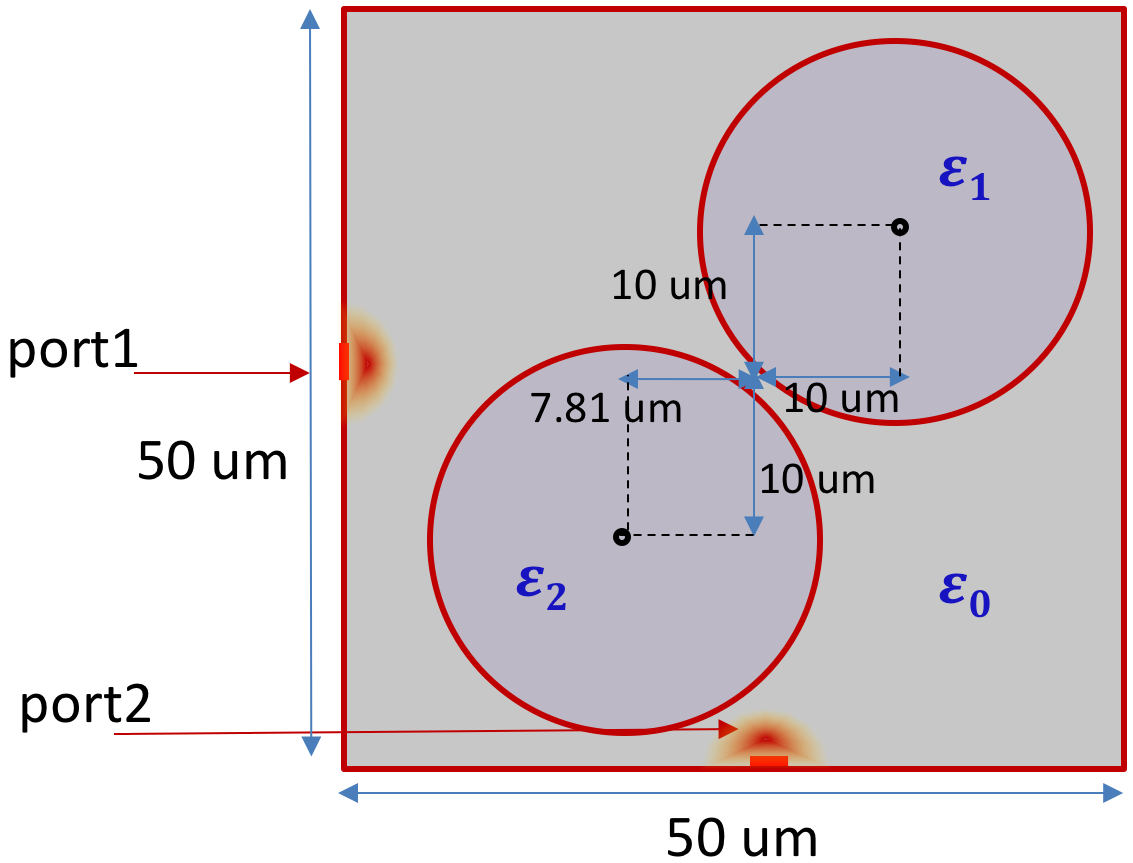}
\caption{Schematic representation of the geometry of the microcavity setup. We report the characteristic lengths used for the scattering analysis reported in Fig. 2 of the main text. The distances are measured from the geometrical center of the cavity.}
\label{FigS1}
\end{figure}

The geometry of our setup is specified in Fig. \ref{FigS1}. The setup consists of a square optical microcavity with side length $50\mu m$.
The cavity is filled by a dielectric medium with a relative permittivity $\epsilon=11.6$ (Silicon) which is perturbed
in two circular regions with radius $12.5\mu$m. We assume that the cavity is surrounded with perfect electric conductors (PECs).

We attach to the structure two ports at the middle of the left and bottom cavity sides, see \ref{FigS1}.
The ports are assumed to be two identical ideal lumped ports with characteristic impedance $Z_0=50$Ohms
with a length of 1/20 of the cavity side. To perform the scattering analysis, we probe the system by illuminating the structure 
by the normal incidence of a plane wave from the left port. The associated scattering parameters $S_{1,1}$ and $S_{2,1}$ result
from a scattering analysis performed using COMSOL Multiphysics software \cite{SCOMSOL}. The transmittance $T=|S_{2,1}|^2$ is reported in Fig. (2) 
of the main text, while the associated reflectance $R=|S_{1,1}|^2=1-|S_{2,1}|^2$.

The calculation of the radiative pumped energy density $Q(\omega)$ in Fig. (3) of the main text, also requires a set of scattering analysis 
for different parameters of the pumping cycle. There, we considered the parametric modulation cycle $\varepsilon_{1(2)}=\varepsilon(U+\Delta\varepsilon_{1(2)})$ where $(\Delta\varepsilon_{1},\Delta\varepsilon_{2})=
\frac{\Delta}{\varepsilon}(-\sin x,\cos x)$ for $x\in[0,2\pi)$ and steps $\Delta x=0.02\pi$. Then, we obtained the scattering 
coefficients for the calculation of $Q(\omega)$, which is calculated using Eqs. (10) and (12) of the main text.

\section*{Appendix A: Useful expressions and properties of the Floquet scattering matrix.}

First, we discuss a few properties of the Floquet scattering matrix that we use to derive our main results.
As we have discussed in the main text, the element of the Floquet scattering matrix ${\bf S}^F_{\alpha,\beta}(\omega,\omega^\prime)$ 
describes the amplitude of the electromagnetic field entering the reservoir $\alpha$ with frequency $\omega$ given that it was 
emitted from reservoir $\beta$ at a frequency $\omega^\prime$. 
Here, the system is periodically modulated with frequency $\Omega$ 
and thus inelastically scatters the field such that $\omega^\prime=\omega+m\Omega=\omega_m$, $m\in \mathbb{Z}$. We assume that 
the the Floquet scattering matrix is unitary, $({\bf S}^F)^\dagger {\bf S}^F={\bf S}^F ({\bf S}^F)^\dagger=I$,  and thus
\begin{equation}
\sum_{\beta,m} |{\bf S}^F_{\alpha,\beta}(\omega,\omega_m)|^2=\sum_{\alpha,m} |{\bf S}^F_{\alpha,\beta}(\omega_m,\omega)|^2=1.
\label{eq:Sunit_1}
\end{equation}
Additionally, there is translational symmetry in the Floquet space, 
$ {\bf S}^F_{\alpha,\beta}(\omega,\omega_m)={\bf S}^F_{\alpha,\beta}(\omega_{-m},\omega)$. 

Next, we provide a list of useful expressions that connect,  in the adiabatic limit, the Floquet scattering matrix with the instantaneous 
scattering matrix. We use the following equations to derive the Eqs. (\ref{SIbias}), (\ref{SI_pump}), and (\ref{SIdis}):
\begin{eqnarray}
  \sum_m |{\bf S}^{F}_{\alpha\beta}(\omega_m,\omega)|^2 &=& \frac{\Omega}{2\pi} \int dt |S^t_{\alpha\beta}(\omega)|^2 
  = \frac{1}{2\pi} \int dx |S^x_{\alpha\beta}(\omega)|^2,  \notag \\
  \sum_{m} |{\bf S}^{F}_{\alpha\beta}(\omega_m,\omega)|^2 \cdot m &=& \frac{\im}{2\pi} \int dt 
  \frac{\partial S^t_{\alpha\beta}(\omega)}{\partial t}  \left( S^t_{\alpha\beta}(\omega)\right)^*
  =\frac{\im}{2\pi} \int dx 
  \frac{\partial S^x_{\alpha\beta}(\omega)}{\partial x}  \left( S^x_{\alpha\beta}(\omega)\right)^* =\frac{A_{\alpha,\beta}(\omega)}{2\pi} , \label{aux}  \\
  \sum_{\beta,m} |{\bf S}^{F}_{\alpha\beta}(\omega_m,\omega)|^2 \cdot m^2 &=& -\frac{1}{2\pi\Omega} \int dt
  \left[ \frac{\partial^2 S^t(\omega)}{\partial t^2}  \left( S^t(\omega)\right)^\dagger \right]_{\alpha,\alpha} 
  = -\frac{1}{2\pi} \int dx
  \left[ \frac{\partial^2 S^x(\omega)}{\partial x^2}  \left( S^x(\omega)\right)^\dagger \right]_{\alpha,\alpha} , \notag
\end{eqnarray}
where $x=\Omega t$, $x\in[0,2\pi)$, and $
A_{\alpha,\beta}=\im \int dt  \frac{\partial S^t_{\alpha,\beta}(\omega)}{\partial t}  \left( S^t_{\alpha,\beta}(\omega)\right)^* $. 
Here, we used the definition of the delta function, $\sum_m e^{\im m \Omega (t-t^\prime)}=\frac{2\pi}{\Omega} \delta(t-t^\prime)$.



\begin{thebibliography}{10}
\bibitem{note1}In our numerical example below, we have used an index variation of the order of 1\%.
This is at the high end of achievable index variations and has been used for a numerical efficiency of the
time-domain simulations with the goal to compare these results with the predictions of the instantaneous 
scattering matrix formalism.
\end{thebibliography}
\end{document}